\documentclass[12pt]{amsart}
\usepackage[T2C]{fontenc}
\usepackage{graphicx}
\usepackage{amssymb}
\usepackage{float}
\usepackage{geometry}
\usepackage{tikz}
\usepackage{float}
\usepackage{url}
\usepackage{mathtools}
\usepackage{amsmath,amsfonts,amsthm}
\usepackage{mathrsfs} 
\usepackage{pgfplots}
\pgfplotsset{compat=1.18}
\usepackage{hyperref}

\newtheorem{definition}{Definition}

\newtheorem{remark}{Remark}

\author{Duaa Abdullah}
\address{\textbf{Duaa Abdullah} 
Physics and Technology School of Applied Mathematics and Informatics,
Moscow Institute of Physics and Technology, 141701, Moscow region, Russia
}
\email{duaa1992abdullah@gmail.com}
\thanks{}

\title{Determining the Qibla Direction by Astronomical and Geometrical Methods}

\date{}

\begin{document}

\begin{abstract}
This paper investigates the determination of the Qibla direction using both astronomical and geometrical approaches. The study reviews historical and classical methods employed by Muslim scholars and astronomers including the use of instruments such as the astrolabe and compass. It further explores spherical trigonometry techniques to precisely calculate the Qibla azimuth from any given location on Earth. The research clarifies geometric constructions and presents a computational model implemented in C++ to facilitate accurate Qibla determination. This interdisciplinary analysis underscores the rich tradition of Islamic astronomy and geometry in solving practical religious requirements, providing both theoretical frameworks and practical algorithms for modern application.
\end{abstract}

\maketitle

\section{Introduction}
Throughout history, islamic geographers and scholars have produced Qibla charts to guide worshippers in aligning themselves toward the Ka'ba~\cite{King1992}. Early approaches, often described as \emph{folk science}, were based on traditional astronomical knowledge from pre-Islamic Arab civilizations. This knowledge, although lacking formal theoretical underpinning or precise calculations, contributed to the development of a sacred geographic worldview. In this worldview, the world was divided into regions radiating from the Ka'ba, with each region's Qibla direction determined by empirical methods rooted in folk astronomy. Subsequently, the introduction of \emph{mathematical science}, largely influenced by Greek scholarship and characterized by theoretical and computational rigor, enabled the formulation of systematic procedures for determining the Qibla from any location.  Despite the varied techniques employed by Muslims over the centuries to face the Ka'ba, it remains unequivocally true that the Qibla occupies a central and profound position in Islamic religious tradition and cultural identity~\cite{Almakky1996}.

The precise dating and original floor plans~\cite{Schumm2020}, including qibla orientations, of many ancient structures remain indeterminate. Scholarly discourse has emerged regarding the proficiency of Muslim architects during the initial two centuries of Islam in precisely determining true qiblas. A portion of the academic community contends that these architects possessed the requisite capability, whereas others dispute this claim. The issue may be nuanced, suggesting that certain architects demonstrated such accuracy while others did not, or that the precision of qibla determination varied temporally and geographically, particularly relative to the distance from qibla targets.

Prayer is an individual duty incumbent on every Muslim man and woman who has reached the age of puberty, is clean and undefiled, possesses a sound mind, has heard the message of the prophet, and is capable of performing it. A child is advised to perform the prayer at the age of seven and instructed to perform it at the age of ten~\cite{Almakky1996}. As stated in the Holy Qur'an, the obligatory prayers should be performed at fixed times. Allah has prescribed that Muslims perform the prayer five times daily.

A recent publication by Nachef and Kadiar~\cite{Abdali1997,Riad1993Samir} posits that the qibla orientation for North America corresponds to the southeast direction. To substantiate this assertion, they cite various classical Islamic jurists. Additionally, their argument is reinforced by testimonials from multiple contemporary Muslim theologians and several scientists from Canada and the United States. The scientists consulted, predominantly geographers, advocate defining the qibla as the rhumb line toward Mecca, which, for the majority of North America, lies within the southeastern quadrant.

Mecca is considered the Qiblah for Muslims from all over the world, and prayer is not valid without facing the Qiblah itself. Due to the importance of determining the Qiblah direction in the lives of Muslims and linking this determination with the astronomical instruments that were developed, perhaps the most important of which are the astrolabe and the quadrant, it was necessary to study the astronomical instruments that contribute to determining the Qiblah direction through different methods in order to establish the direction of the Qiblah, which was formerly called the Qiblah azimuth.

\begin{definition}[\textbf{Qiblah Azimuth}]
It is the intersection point between the horizon circle and a great circle passing through the azimuths of our heads and the heads of the people of Mecca, according to what is mentioned in the books of astronomers.
\end{definition}

There is no doubt that astronomy was known thousands of years before Christ. This is evidenced by what the Babylonians, the ancient Egyptians, and others have left in their records of certain astronomical phenomena, or which were built based on precise astronomical observations. The ancient Babylonians divided their agricultural year into three seasons and took the time when the star Sirius (Al-Shi'ra Al-Yamaniyah) was at a specific position in the eastern sky as the beginning of that year.

Determining the Qiblah means facing the House of Allah Al-Haram from any point on the surface of the Earth. As is known, the Qiblah of Muslims is Mecca. In this research, we will present methods of finding the Qiblah both astronomically and geometrically, through the study of the Qiblah azimuth and its clear determination. We will try to clarify a certain model to determine the azimuth using geometric laws which are indispensable for us. Then, we will present a mechanism for determining the Qiblah direction according to certain rules and steps, using astronomical instruments made for this purpose, along with other geometric and astronomical matters.

\subsection{Research Objective}
For over 1,400 years, the obligation to pray and to perform various ritual acts in a sacred direction toward the Ka'ba has been of paramount importance to Muslims all over the world. The fundamental importance within Islam of the concept of Qibla led to the development of Qibla charts, maps, instruments, and related cartographic methods. Many studies\footnote{These studies were discussed through the PhD thesis~\cite{Jasem3}.} have been presented on determining the Qiblah direction in the modern era, but the real importance is in identifying the methods used by the ancients to determine the Qiblah direction. Since many astronomical instruments existed and inevitably performed similar tasks, they differed from each other in the special functions that each instrument performed individually. Through this research, we will provide answers to some important questions, which are:
\begin{enumerate}
    \item How can the Qiblah direction be correctly determined?
    \item What is the margin of error in determining the Qiblah direction, and how to correct it?
    \item What are the most important astronomical instruments used to determine the Qiblah direction?
    \item What geometric laws help us in determining the Qiblah direction?
\end{enumerate}

\section{Qibla Direction Using Astronomical and Geometrical Methods}
The calculation of the azimuth, or directional bearing, between two geographic points relies fundamentally on two critical aspects. The first essential element is the precise determination of the coordinates of these locations, which is typically expressed in terms of their latitudinal and longitudinal values. The second factor involves specifying the nature of the route connecting the two points. Generally, this route is defined either as a great circle path, also known as a geodesic, or alternatively, as a rhumb line. Both aspects introduce distinct implications for the calculation process.
This refers to orienting towards the Qibla by using drawings and geometric shapes that help in determining the Qibla direction. 
Below, we will review some geometrical methods to accurately extract the Qibla direction, and the methods are as follows: 

\subsection{First Method: Determining the Qibla Based on the Compass}
This method relies on what David King mentioned in his research~\cite{KINGCVDS}, which provided an explanation on determining the Qibla direction, and upon which some parts of this method we present are based.

This problem has attracted considerable scholarly attention from several prominent Muslim scientists, such as al-Khwarizmi\footnote{\textbf{\textit{Al-Khwarizmi’s}} greatly influential treatise on algebra gave the name to that discipline. 
The word algorithm derived from his name has become a household word with the spread of computers.} (780–850), al-Battani\footnote{\textbf{\textit{Al-Battani}} determined several astronomical quantities with remarkable precision.} (858–929), Abu al-Wafa al-Buzjani\footnote{\textbf{\textit{Abu al-Wafa}} made significant contributions to mathematics and astronomy. 
He interrelated the six main trigonometric functions. 
He is, arguably, credited with discovering a component of the moon’s motion, rediscovered 600 years later by Tycho Brahe.} (940–997), Ibn al-Haitham\footnote{\textbf{\textit{Ibn al-Haitham}}, famous for his foundational work on optics, also worked prolifically in mathematics and astronomy. 
He was a pioneer of the scientific method consisting of observation, hypothesis formulation, deduction, and experimental verification.} (965–1040), al-Biruni\footnote{\textbf{\textit{Al-Biruni}} is described by Durant~\cite{Durant1950}.} (973–1048), and al-Tusi\footnote{
\textbf{\textit{Al-Tusi}} systematized trigonometry (plane and spherical) as a discipline independent of astronomy. 
He also formulated a non-Ptolemaic model of planetary motion based on spheres.} (1201–1274). 
These figures represent some of the most influential contributors to medieval scientific knowledge. 
Additionally, significant original advancements in the determination of the qibla were made by numerous scholars, including Habash al-Hasib\footnote{\textbf{\textit{Habash al-Hasib}} introduced new trigonometric functions.}
(circa 850), al-Nayrizi (circa 897), Ibn Yunus\footnote{\textbf{\textit{Ibn Yunus}} compiled very accurate astronomical tables, studied the motion of the pendulum leading to the invention of mechanical clocks, and invented several astronomical instruments.} 
(circa 985), ibn al-Banna al-Marrakashi\footnote{\textbf{\textit{Al-Banna}} did novel work on continued fractions, power sums, and what can be now be recognized as binomial coefficients.}
(1256–1321), al-Khalili (circa 1365), and Ibn al-Shatir\footnote{\textbf{\textit{Ibn al-Shatir}} constructed a model of planetary movements.} (1306–1375). 
Although these latter figures are often less widely recognized, they nonetheless made substantial contributions to the fields of astronomy and mathematics\footnote{\textbf{Al-Battani’s} construction, 
described in~\cite{Bannuri1939,King1993,Ludhianavi1970}.}.
\begin{enumerate}
    \item Draw a circle with a fairly large radius, similar to the circle from which we extract the meridians, and divide it into four sections by two perpendicular lines.
    \item Identify the four cardinal directions on the circle.
    \item We then have four equal sections; we divide each quarter of the circle into ninety equal parts. Divide the line that goes from the north point of the circle's circumference to the center of the circle on a half-line — the line from the north point to the center — into ninety equal parts.
    \item Determine the longitude of the city whose Qibla direction we want to determine, say \(s\).
    \item Take the city’s longitude as the number marked on the circle from the east point to the center of the circle, which is from the south line to the west, lying on the equator line from east to west.
    \item In this way, we have determined the meridians of that city whose Qibla direction we want to determine, then square the circle with two lines from that marker similar to the first two lines, and leave these first two lines as we no longer need them.
    \item Label the ends of the lines with the directions east, west, north, and south, and make these lines clearly visible on their edges.
    \item Determine the longitude and latitude of Mecca and the longitude and latitude of the city we are in.
    \item Take from the east line on the circle the amount corresponding to Mecca’s longitude—if taking length—and the amount corresponding to the latitude—if taking latitude. In fact, this can be clarified through Figure~\ref{fign1}, where we illustrate the mechanism of determining the Qibla.
\begin{figure}[H]
\begin{center}
  \includegraphics[width=.4\linewidth]{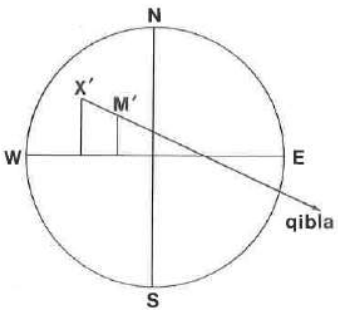}
\caption{Dividing the circle and placing the quarters and directions.}~\label{fign1}
\end{center}
\end{figure}
    \item Then take from the lines on the edge of the circle the segments corresponding to Mecca’s longitude; if the length is completed, draw a line from that position from the circle's point on the line following south from the circle to its center.
    \item Use the compass; place one end on Mecca’s latitude position and the other end at the center of the circle. Draw a new circle inside the large circle, and wherever this new circle intersects Mecca’s longitude line, mark it; this indicates Mecca’s position on Earth, as shown in Figure~\ref{fign2}.
\begin{figure}[H]
\begin{center}
  \includegraphics[width=.4\linewidth]{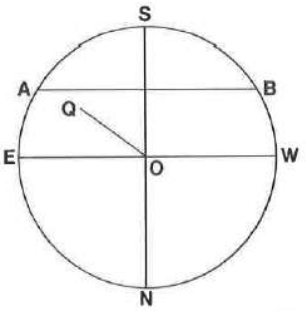}
\caption{The new circle we obtained inside the large circle.}~\label{fign2}
\end{center}
\end{figure}
\end{enumerate}

Then go to the city we are in from the east or west side and mark its longitude and latitude as we did for Mecca to know its position on Earth. Then, place one end of the compass on Mecca’s position and the other end on the city’s position and draw a circle with Mecca at its center and the city on its circumference. Square the circle with two lines: one from the city we calculate for to Mecca and to what corresponds on the other side of the circle; this is the Qibla direction, and the other line is by squaring the Qibla where it falls, God willing.

\subsection{Second Method: Determining the Qibla Using the Astrolabe}
Astronomy is defined as the science that studies the movement of celestial bodies and heavenly objects, requiring different instruments to observe the movements of planets and nearly fixed stars relative to the observer. These observing tools may vary in type and design, yet they ultimately serve as means of studying astronomy and uncovering its amazing secrets and truths. Among these observation tools are the "astrolabe," the responding windcatcher, the ringed one, and others. The astrolabe is defined as: a device used to measure the altitude of stars and calculate their dimensions and positions in the celestial sphere~\cite{khaz001}.

To determine the Qibla direction, we need to know Mecca’s longitude and latitude, as illustrated in Figure~\ref{fign1}, showing the Qibla direction.

\begin{definition}[\textbf{Equator Line}~\cite{GHRTn1}]
It is a circle of latitude around the Earth, equidistant from the poles, dividing the Earth into two halves: northern and southern. The length of day and night is equal at it. "The tropical regions: the land area located between the two tropics."
\end{definition}

\subsubsection{Determining the Qibla Direction Using the Astrolabe}
To determine the Qibla direction using the astrolabe, according to a manuscript message about working with the astrolabe by Umayyah ibn Abu Al-Salt~\cite{Jasem2}, the author explains how to do this by saying: "If you want to do so: extract the meridian line, the equator line, and the points of east, west, north, and true south. Place the astrolabe on the guide that determines these directions. This point is specified on the face that the preceding chapter provided, and it is completely clear to you in which quarter of the horizon the city lies whose direction you want to know. You already know this from its longitude and latitude, because if they agree in longitude and your city’s latitude is greater than the other city’s latitude, (the other city from your city) aligns with the midpoint of the south; if the latitude of your city is less, that city aligns with the midpoint of the north; if it is greater, it is west of it. Once you know the quarter where the desired city lies, take from one of the two points determining it the number of parts of the named distance between that point and the city. Mark the alidade letter on it; by doing so you determine the alidade letter pointing to that city whose location is its solution, which is the Qibla, or otherwise. As for knowing the distances between cities in azimuth parts, this is not possible to mention in this book as it is another craft other than working with the astrolabe, which is the craft of geometry. Instead, we take it here from those who have calculated it using the known tool called the Cutter for a city with this city and placing it for those who need it. The Qibla azimuth can be known by many means, but it is a special one for the astrolabe."

\paragraph{Practical Application:}
We must then follow the following steps:
 \begin{enumerate}
     \item First: Extract the meridian line, equator line, points of east, west, north, and south, then
     \begin{enumerate}
         \item Place the astrolabe on the location determining these directions, and you must know which quarter of the horizon the city lies in whose direction you want to know because you know this from its longitude and latitude.
         \item If the longitudes agree and the latitude of your city is greater than the other city, (the other city being from your city) aligns with the midpoint of the south.
     \end{enumerate}
     \item Second: Determine the Qibla direction through the city's latitude as follows:
     \begin{enumerate}
         \item If your city’s latitude is less, that city aligns with the midpoint of the north.
         \item Once you know the quarter where the desired city lies, take from one of the two points determining it the number of parts of the named distance between that point and the city.
         \item Then place the alidade letter on it, and by doing so you mark the alidade letter that points to the desired city whose solution is the Qibla.
     \end{enumerate}
    \item Third: 
    Compare the longitudes and latitudes of the two cities — the city you are in and the city whose Qibla direction you want to determine. If the longitudes agree, compare the latitudes. If your city’s latitude is greater than the other city’s latitude, the other city aligns with the midpoint of the south. Then determine the quarter in which the city lies on the horizon. Take from one of the two points determining it — eastern or western point — the number of parts of the named distance between those points and that city as specified on the city's plate. Mark the alidade letter on the number of distance parts; thus you fix the alidade letter pointing to the desired city whose solution is the Qibla.
    \item Fourth: If your city’s latitude is less than the other city’s latitude, the city aligns with the midpoint of the north. Then determine the quarter in which the city lies on the horizon. Take from one of the two points determining it — eastern or western point — the number of parts of the named distance between that point and the city, as indicated on the city's plate. Mark the alidade letter on the number of distance parts; then mark this alidade letter pointing to the city whose solution is the Qibla.
 \end{enumerate}

From the above study to determine the Qibla azimuth, we can say that the north and south points refer to the Tropics of Cancer and Capricorn. Once we know that, we determine the longitudes and latitudes for the city whose Qibla we want to find and for the city we are in. We place the astrolabe in the place that shows these directions, which are the points of east, west, north, and true south. According to our study, we present the meridian line through definition~\ref{deffn3}, as it is linked to the Qibla azimuth and the equator line.

\begin{figure}[H]
\begin{center}
  \includegraphics[width=.6\linewidth]{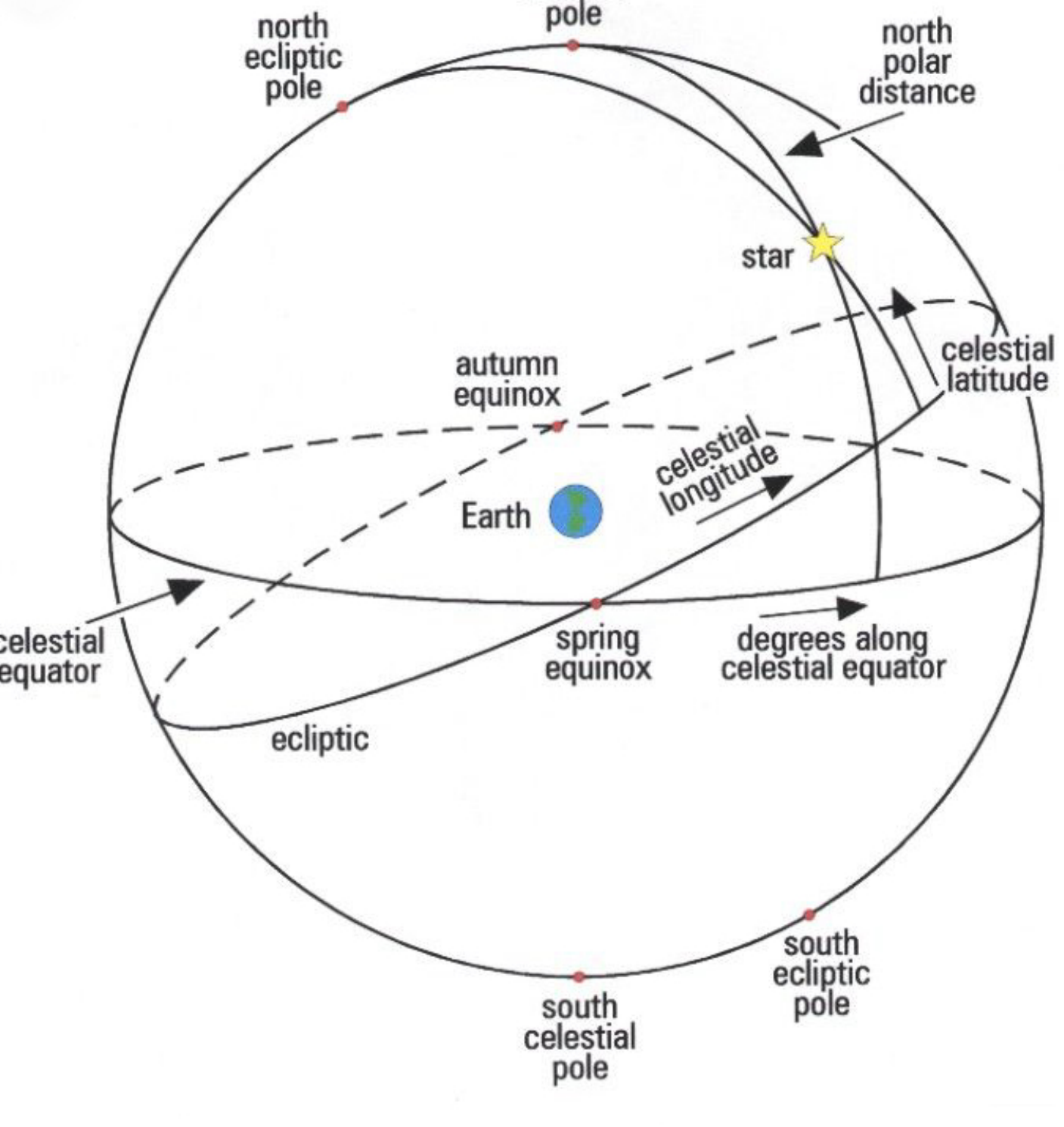}
\caption{Image depicting the position of the Tropics of Cancer and Capricorn and the connecting line between them, which is part of the zodiac circle’s path.}~\label{fign3}
\end{center}
\end{figure}

\begin{definition}[\textbf{Meridian Line}]~\label{deffn3}
It is the line representing the highest altitude of the sun at noon and acts as the middle line between the eastern and western halves, passing through the poles of the first and sixth houses, intersecting them at the points of south and true north, connected by the "zenith" line. It also passes through the second and fourth houses, which represent the earth’s and sky’s pegs; that is, the line connecting the Tropic of Cancer and Tropic of Capricorn.
\end{definition}

\begin{figure}[H]
\begin{center}
  \includegraphics[width=.5\linewidth]{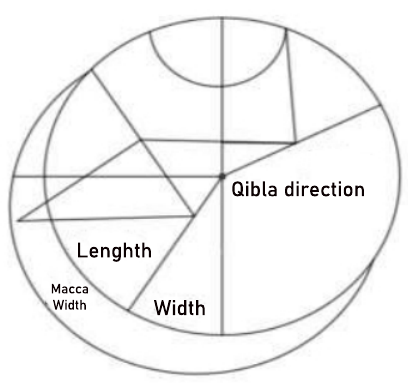}
\caption{Image showing the Qibla direction relative to Mecca.}~\label{fign4}
\end{center}
\end{figure}

Other astronomical instrument methods are similar and do not differ much from this method, regardless of the specific astronomical device used. The question now is how can the Qibla direction be determined by astronomical calculations? Of course, this involves another topic: spherical triangles, as astronomy relies on spherical triangles as much as on plane triangles, since a triangle drawn on a page is like a triangle drawn on the Earth or celestial sphere.

The sides of a spherical triangle are arcs forming the spherical triangle, so the side length of a spherical triangle as an angle in degrees times the radius of the sphere on which the triangle is drawn equals the arc length.

To determine the Qibla direction for a particular city, we follow the following rule:

\begin{enumerate}
    \item Determine the longitude of the city you are in as latitude, but let \(m\) be the longitude and \(n\) the latitude.
    \item Determine the longitude of Mecca as latitude, but let \(a\) be the longitude and \(b\) the latitude.
    \item Take the difference in longitude between the city you are in and Mecca defined by:
\[
x = 90 - a \quad y = 90 - m
\]
    \item Take the difference in latitudes, latitude of Mecca and the city you are in, defined by:
\[
z = n - b
\]
    \item Substitute in the triangle formulas which are:
\begin{equation}~\label{eq11}
\tan\left(\frac{A + B}{2}\right) = \frac{\cos\left(\frac{x - y}{2}\right) \cot \frac{z}{2}}{\cos \left(\frac{x + y}{2}\right)} \quad , \quad \tan\left(\frac{A - B}{2}\right) = \frac{\sin \left(\frac{x - y}{2}\right) \cot \frac{z}{2}}{\sin \left(\frac{x + y}{2}\right)}
\end{equation}
By solving these equations, we obtain the angle \(A\), and thus the Qibla direction for the city we are in is:
\[
r = 180 - A
\]
\end{enumerate}

Let us take an example to determine the Qibla direction for the city of \textbf{Aleppo}:

The longitude and latitude are: \(n = 37.16, \quad m = 36.18\).

Mecca’s longitude and latitude are: \(b = 39.50, \quad a = 21.25\).

Now, determine the coordinates: \(x = 90 - 21.25 = 68.75, \quad y = 90 - 36.18 = 53.82\).

Calculate the latitude difference and get: \(z = 39.50 - 37.16 = 2.34\).

Thus, we have the relation:
\[
\tan\left(\frac{A+B}{2}\right) = \frac{\cos(7.4650)\cot(1.17)}{\cos(61.2850)},
\quad
\tan\left(\frac{A-B}{2}\right) = \frac{\sin(7.4650)\cot(1.17)}{\sin(61.2850)}.
\]

From this relation, we get
\[
\tan\left(\frac{A+B}{2}\right) = 101.0482, \quad \tan\left(\frac{A-B}{2}\right) = 7.2534.
\]

The following figure shows the Qibla direction from the city of Aleppo to Mecca.
\begin{figure}[H]
\begin{center}
\includegraphics[width=.5\linewidth]{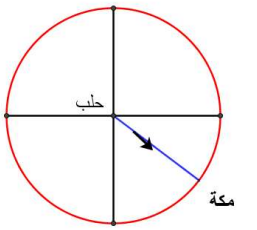}
\caption{Image showing the Qibla direction relative to Aleppo city towards Mecca.}~\label{fign5}
\end{center}
\end{figure}

\section{Basic Spherical Trigonometry Equations for Qibla Direction}

The Qibla is the direction from any point on Earth toward the Kaaba in Mecca. This is a classic problem in \emph{spherical astronomy}, solved using the \emph{spherical law of cosines} on the unit sphere (Earth modeled as a sphere). Let $P$: Observer's position with latitude $\phi_P$ (positive north), longitude $\lambda_P$, $K$: Kaaba (Mecca) at latitude $\phi_K = 21.4225^\circ$ N, longitude $\lambda_K = 39.8262^\circ$ E (approximate fixed values), $\Delta\lambda = \lambda_K - \lambda_P$: Longitude difference and $q$: Qibla bearing (azimuth) measured \emph{clockwise from local north} ($0^\circ$ = north, $90^\circ$ = east, etc.) The \emph{exact formula} derived from the spherical cosine law is:
\begin{equation}~\label{eqqq1ssd1}
\tan q = \frac{\sin \Delta\lambda}{\cos \phi_P \tan \phi_K - \sin \phi_P \cos \Delta\lambda}.
\end{equation}

Thus, the Qibla direction is:
\begin{equation}~\label{eqqq1ssd2}
q = \left(\tan^{-1}\left(\frac{\sin \Delta\lambda}{\cos \phi_P \tan \phi_K - \sin \phi_P \cos \Delta\lambda} \right)\right)^2.
\end{equation}

\begin{remark}
Use the two-argument arctangent $(\tan^{-1}(y, x))^2$ to get the correct quadrant ($q \in [0^\circ, 360^\circ)$).
\end{remark}

Represent points as unit vectors in 3D Cartesian coordinates (ECEF frame, z-axis through north pole):
\begin{equation}~\label{eqqq1ssd3}
   \vec{P} = \begin{pmatrix} \cos \phi_P \cos \lambda_P \\ \cos \phi_P \sin \lambda_P \\ \sin \phi_P \end{pmatrix}, \quad
    \vec{K} = \begin{pmatrix} \cos \phi_K \cos \lambda_K \\ \cos \phi_K \sin \lambda_K \\ \sin \phi_K \end{pmatrix}.
\end{equation}
The \emph{great circle} from $P$ to $K$ lies in the plane defined by $\vec{P}$, $\vec{K}$, and origin.  The \emph{local north direction} at $P$ is the unit vector in the meridian plane: 
\begin{equation}~\label{eqqq1ssd4}
\vec{N} = \begin{pmatrix} -\sin \phi_P \cos \lambda_P \\ -\sin \phi_P \sin \lambda_P \\ \cos \phi_P \end{pmatrix}.
\end{equation}
The \emph{east direction} at $P$ is perpendicular:  
\begin{equation}~\label{eqqq1ssd5}
  \vec{E} = \begin{pmatrix} -\sin \lambda_P \\ \cos \lambda_P \\ 0 \end{pmatrix}.
\end{equation}
 The direction vector from $P$ to $K$ along the great circle is the component of $\vec{K} - \vec{P}$ orthogonal to $\vec{P}$ as $ \vec{D} = \vec{K} - (\vec{K} \cdot \vec{P}) \vec{P}.$  Project $\vec{D}$ onto local north and east as $ q = \arctan2(\vec{D} \cdot \vec{E},\ \vec{D} \cdot \vec{N}).$  After algebraic expansion according to~\eqref{eqqq1ssd2}--\eqref{eqqq1ssd5} (using $\vec{K} \cdot \vec{P} = \cos c$, where $c$ is angular distance), this simplifies exactly to the tangent formula above.

This is the \emph{standard exact method} used in astronomy and Islamic geography (e.g., in prayer apps, NOAA solar calculator adaptations). We presented the Qibla direction among Table~\ref{tab01Qibla} for some capital.  Only \emph{algebraic spherical trigonometry} is needed. No integrals or differential equations are involved in determining the Qibla direction.

\begin{table}[H]
    \centering
    \begin{tabular}{|l|c|c||l|c|c||l|c|c|}
    \hline
 Damascus            & 171.09 & S & Kabul & 250.77 & W & Tiran & 133.61 & SE \\ \hline
 Algiers & 105.39 & E & Luanda & 40.39  & NE & Andorra la Vella & 130.02 & SE\\ \hline
 Yerevan & 193.35 & S & Baku  & 207.18 & SW & Manama  & 246.23 & W \\ \hline
Nassau  & 58.33  & NE & Dhaka  & 277.57 & W & Minsk  & 159.17 & S \\ \hline Brussels & 123.48 & SE & Thimphu & 273.78 & W  & Sarajevo  & 134.77 & SE \\ \hline
Brasília & 68.76  & E& Sofia  & 141.86 & SE & Ouagadougou  & 71.39  & E \\ \hline
Bujumbura & 22.01  & N     & Yaoundé & 54.99  & NE  & Praia& 73.57  & E     \\ \hline
N'Djamena & 65.30  & NE    & Beijing & 278.89 & W & Kinshasa & 42.00  & NE   \\ \hline
Brazzaville & 42.20  & NE & Yamoussoukro & 66.66  & NE & Havana              & 56.07  & NE \\ \hline
Djibouti  & 342.54 & N     & Roseau              & 66.34  & NE    & Santo Domingo & 63.01  & NE \\ \hline
Asmara  & 7.84   & N & Addis Ababa & 4.67   & N & Helsinki & 158.23 & S      \\ \hline
Khartoum & 34.05  & NE    & Mbabane & 13.35  & N & Stockholm  & 138.24 & SE\\ \hline
\end{tabular}
    \caption{Lists the Qibla direction (bearing from true north, clockwise)}
    \label{tab01Qibla}
\end{table}
\section{Qibla Determination Application}
In the first Section, we discussed methods of determining the Qibla through astronomical and geometric approaches, and we calculated the Qibla azimuth for the city of Aleppo with precision. In this chapter, we will demonstrate how we can apply Qibla determination; that is, modeling this application in general to obtain the Qibla direction by the fastest and best possible methods.

First, we will use programming in C++ because it is the best and easiest in terms of capabilities; thus, we can easily implement this on a mobile phone. We will not delve deeply into this language since it is not the subject of our research. The first program will be designed to obtain the coordinates x, y, z, and the second phase involves solving spherical equations for which we use the Wolfram Alpha program. In this way, we have solved the problem of modeling the Qibla azimuth and presented it in the simplest possible form. The program is as follows, and I will not mention the meaning of the symbols since I have explained them on the previous page and will adhere to them.

\begin{verbatim}
#include <iostream>
#include <math.h>
using namespace std;
int main ()
{
float m,n,a,b,x,y,z,s,d,r,e;
cout << "pleas enter m= " ;
cin >> m;
cout << " \n pleas enter n= " ;
cin >> n;
cout << "\n pleas enter a= " ;
cin >> a;
cout << "pleas enter b= " ;
cin >> b;
x=90-a;
y=90-m;
cout << "we found x=" << x <<"and y= " << y<< endl;
z=b-n;
cout << " the change in width is z=" <<z << endl;
cout <<"************" << endl;
s= cos ((x-y)/2)* (cos(z)/ sin (z));
cout << "x-y /2 =" << ((x-y)/2) << endl;
d= cos ((x+y)/2);
cout << " s= " << s << " d= " << d<< endl;
cout << "tan ((A+B)/2) = " << s/d << endl;
r= sin ((x-y)/2)* (cos(z)/ sin (z));
e= sin ((x+y)/2);
cout << " r=" << r << " e= " << e << endl;
cout << "tan ((A-B)/2) = " << r/e<< endl;
cout <<"***********" << endl;
return 0;
}
\end{verbatim}

 This program does not precisely find the azimuth of the region for which the azimuth is to be found, but it leads us to the relationships that rely in their solution on spherical triangles, which in turn give us the specific azimuth of the country for which we want to calculate the azimuth.
\section{Calculate the Great-Circle Path}
Computers allow the calculation of the Qibla to considerable accuracy from any point in the world as long as the latitudes, longitudes, and reference ellipsoid are accurately supplied.

The following formula gives the Qibla (azimuth to Makkah) based on the Earth as a sphere:
\begin{equation}
\mathrm{Az}=\frac{\arctan_{2}{\cos \Phi_{1} \sin (\lambda_{1}-\lambda_{2})}}{\cos \Phi_{2} \sin \Phi_{1}-\sin \Phi_{2} \cos \Phi_{1} \cos (\lambda_{1}-\lambda_{2})}
\end{equation}

where $\Phi_{1}, \lambda_{1}=$ latitude (+ if north, - if south) and longitude (+ if east, - if west of Greenwich), respectively, of the Ka'ba in Makkah, $\Phi_{2}, \lambda_{2}=$ latitude and longitude, respectively, of the worshipper's location, $\mathrm{Az}=$ Azimuth or Qibla, measured clockwise from north $\arctan _{2}=\arctan$ function adjusted for quadrant; same as ATAN2 or DATAN2 function in FORTRAN, arranged so that if the denominator in this formula is negative, $\pi$ radians $\left(180^{\circ}\right)$ is added to Az (and $2 \pi$ is then subtracted if Az exceeds $\pi$ )

In Washington, D.C., the precise orientation for a worshipper to face is approximately $56^{\circ} 30^{\prime}$ measured clockwise from true north. This corresponds to a bearing roughly 11 degrees to the east of the northeast direction. If one were to mistakenly use the rhumb line for this calculation, the resulting direction would be near $99^{\circ} 50^{\prime}$ clockwise from true north, placing it about 10 degrees south of due east. These angular measurements are all referenced to the GRS-80 ellipsoid. 

\section{ Conclusion}

 In summary, we can say through the research we conducted that the direction of the Qibla can be determined in more than one way, including mathematical methods, astronomical methods, and methods based on calculations related to spherical triangles. The Arabs excelled in this as much as possible and were able to find the azimuth by various methods, using many astronomical instruments, the most important of which is the astrolabe. We were able to develop an approximate program that helps us shorten many relationships, representing a clear effort for the researcher.

\end{document}